\documentclass{iaus}
\usepackage{graphicx}

\title[JD 11.~~First stars and cosmic infrared background] %% give here short title %%
{Imprint of first stars era in the cosmic infrared backround
fluctuations }

\author[A. Kashlinsky]   %% give here short author list %%
{A. Kashlinsky
}

\affiliation{SSAI and Observational Cosmology Lab, Code 665,
Goddard Space Flight Center, Greenbelt, MD 20771, U.S.A. \\email:
{\tt Alexander.Kashlinsky.1@nasa.gov}}

\pubyear{2008}
\volume{250}  %% insert here IAU Symposium No.
\pagerange{xxx--xxx}
 \jname{Massive stars as cosmic engines}
\editors{Fabio Bresolin, Paul Crowther \& Joachim Puls, eds.}
\begin{document}

\maketitle

\begin{abstract}
We present the latest results on CIB fluctuations from early
epochs from deep Spitzer data. The results show the existence of
significant CIB fluctuations at the IRAC wavelengths (3.6 to 8
$\mu$m) which remain after removing galaxies down to very faint
levels. These fluctuations must arise from populations with a
significant clustering component, but only low levels of the shot
noise. There are no correlations between the source-subtracted
IRAC maps and the corresponding fields observed with the {\it HST}
ACS at optical wavelengths. Taken together, these data imply that
1) the sources producing the CIB fluctuations are individually
faint with $S_\nu<$ a few nJy at 3.6 and 4.5 $\mu$m; 2) have
different clustering pattern than the more recent galaxy
populations; 3) are located within the first 0.7 Gyr (unless these
fluctuations can somehow be produced by - so far unobserved -
local galaxies of extremely low luminosity and with the unusual
for local populations clustering pattern), 4) produce contribution
to the net CIB flux of at least 1-2 nW/m$^2$/sr at 3.6 and 4.5
$\mu$m and must have mass-to-light ratio significantly below the
present-day populations, and 5) they have angular density of
$\sim$ a few per arcsec$^2$ and are in the confusion of the
present day instruments, but can be individually observable with
{\it JWST}. 
\end{abstract}

\firstsection % if your document starts with a section,
\section{Introduction}

The cosmic infrared background (CIB) is a repository of emissions
throughout the entire history of the Universe. The recent years
have seen significant progress in CIB studies, both in identifying
and/or constraining its mean level (isotropic component) and
fluctuations (see Kashlinsky 2005 for a recent review). The CIB
contains emissions also from objects inaccessible to current (or
even future) telescopic studies and can, therefore, provide unique
information on the history of the Universe at very early times.
One particularly important example of such objects, of particular
reference to this conference, concerns Population III stars
(hereafter Pop~III), the still elusive zero-metallicity stars
expected to have been much more massive than the present stellar
populations (see Bromm \& Larson 2004 for a recent review).
Herebelow I will use the term "era of the first stars", or
"Pop~III era", with the understanding that the actual era may be
composed of objects of various nature from purely zero-metallicity
stars, to low- metallicity stars to even possibly mini-quasars
whose contribution to the CIB is driven by energy released by
gravitational accretion, as opposed to stellar nucleosynthesis.

Extensive numerical investigations of collapse and fragmentation
of the first objects forming out of density fluctuations specified
by the standard $\Lambda$CDM model suggest that Pop III stars were
quite massive and lived at $z> 10$, well within the first Gyr of
the Universe's evolution. If predominantly massive, they are
expected to have left a significant level of diffuse radiation
redshifted today into the IR, and it has been suggested that the
CIB contains a detectable contribution from Pop~III in the
near-IR, manifest in both its mean level and its anisotropies
(e.g. Bond et al 1986, Santos et al 2002, Salvaterra \& Ferrara
2003, Cooray et al 2004, Kashlinsky et al 2004).

In the past several years a group of us (Kashlinsky, Arendt,
Mather \& Moseley 2005, 2007a,b,c - hereafter KAMM1, KAMM2, KAMM3,
KAMM4) have used deep-integration {\it Spitzer} data to measure
the CIB fluctuations component arising from early populations.
These provide first observational insights into the global
evolution of the Universe at early cosmic epochs. Our measurements
revealed significant CIB fluctuations at the IRAC wavelengths (3.6
to 8 $\mu$m) which remain after removing galaxies down to very
faint levels (KAMM1, KAMM2). These fluctuations must arise from
populations that have a significant clustering component, but only
low levels of the shot noise (KAMM3). Furthermore, there are no
correlations between the source-subtracted IRAC maps and the
corresponding fields observed with the {\it HST} ACS at optical
wavelengths (KAMM4). Taken together, these data imply that 1) the
sources producing the CIB fluctuations have a very different
clustering pattern than galaxies at intermediate redshifts and are
individually faint with $S_\nu<$ a few nJy at 3.6 and 4.5 $\mu$m;
2) are located within the first $\simeq 0.7$ Gyr (unless these
fluctuations can somehow be produced by - so far unobserved -
local galaxies of extremely low luminosity and with the unusual
for local populations clustering pattern), 3) they produce
contribution to the net CIB flux of at least 1-2 nW/m$^2$/sr at
3.6 and 4.5 $\mu$m and must have mass-to-light ratio significantly
below the present-day populations, and 4) their angular density is
$\sim$(a few) arcsec$^{-2}$, so they are in the confusion of the
current instruments, but can be individually observable with {\it
JWST}.

Below, I will discuss the latest measurements of the fluctuations
in the CIB by our team (Kashlinsky, Arendt, Mather \& Moseley -
KAMM) and explore their implications for the nature of the sources
contributing to these anisotrpies, specifically in the era of the
first stars. Following the Introduction, Sec. 2 reviews the
current measurements at both near-IR (IRAC) and optical (ACS)
wavelengths and Sec. 3 discusses the nature of the cosmological
populations producing these CIB anisotropies.

\section{Source subtracted CIB fluctuations vs optical galaxies}

Before we discuss the interpretation of the KAMM measurements, it
is important to review the steps done in the analysis leading to
the measured CIB fluctuations. The data have been assembled from
the individual AORs using the self-calibration method from
\cite{fixsen}. The exposure times ranged from $\sim 8-9$
hours/pixel for the initial $5^\prime\times 10^\prime$ QSO 1700
field (KAMM1) to $\sim 23-25$ hours/pixel in each of the four
GOODS fields of $10^\prime \times 15^\prime$ (KAMM2) The latter
have been observed at two different epochs separated by $\sim 6$
months allowing us to better handle zodiacal gradients and
possible instrument systematics. The images have been cleaned of
foreground galaxies and stars in two steps: 1) all pixels with
flux fluctuations exceeding a fixed number of standard deviations
of the image were iteratively blanked, and 2) the residual faint
parts of the sources were removed iteratively using a modified
CLEAN algorithm (Hogbom 1974) where the maximum pixel intensity is
located and the wide PSF is then scaled to half of this intensity
and subtracted from the image. In the first step, which defines
the final mask, it is important to remain in the regime when the
fraction of the removed pixels is small enough to allow a reliable
computation of the power spectrum using FFTs. The second step
allows us removal of progressively fainter foreground populations
and enables a better characterization of the remaining (and
removed) populations.

\begin{figure}[ht]
% \vspace*{-2.0 cm}
\begin{center}
 \includegraphics[width=5in]{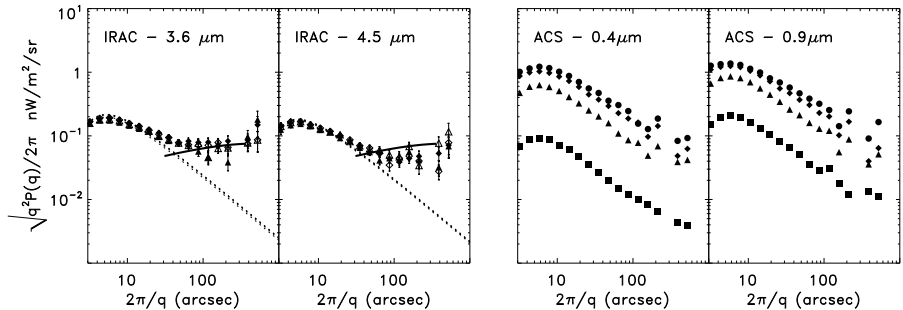}
% \vspace*{-1.0 cm}
 \caption{{\it Left}: Source-subtracted CIB
fluctuations from KAMM2 at 3.6 and 4.5 $\mu$m. Four sets of
symbols correspond to the four GOODS fields. Dotted lines show the
shot-noise contribution. Solid line shows the slope of sources at
high-$z$ with the $\Lambda$CDM model spectrum of the same
amplitude at 3.6 and 4.5 $\mu$m. {\it Right}: CIB fluctuations due
to ACS galaxies for the $972\times 972$ 0.6$^{\prime\prime}$ pixel
field at HDFN-Epoch2 region for the ACS B and z-bands. Filled
circles correspond to ACS galaxies fainter than $m_0=21$ with the
mask defined by the clipping. Filled diamonds, triangles and
squares correspond to fluctuations produced by sources fainter
than $m_0+2,m_0+ 4, m_0+6$.}
   \label{fig1}
\end{center}
\end{figure}

Left panels of Fig. \ref{fig1} show the source-subtracted CIB
fluctuations (the instrument noise has been subtracted) for four
GOODS fields at 3.6 and 4.5 $\mu$m adopted from KAMM2. The
detected signal is significantly higher than the instrument noise
and the various systematics effects cannot account for it.
Similarly, fluctuations due to emissions in the Solar System and
the Galaxy are too weak, except at 8  $\mu$m where Galactic cirrus
may contribute to the measured signal. There was a statistically
significant correlation between the channels for the regions of
overlap meaning that the same population is responsible for the
fluctuations. The correlation function at deeper clipping cuts,
when too few pixels were left for Fourier analysis, remains the
same and is consistent with the power spectrum numbers (KAMM1).
The signal is to a good accuracy isotropic on the sky, as required
by its extragalactic origin, and must thus contain contributions
from the ``ordinary" galaxies and from unresolved populations at
high $z$.

\begin{figure}[ht]
% \vspace*{-2.0 cm}
\begin{center}
 \includegraphics[width=5in]{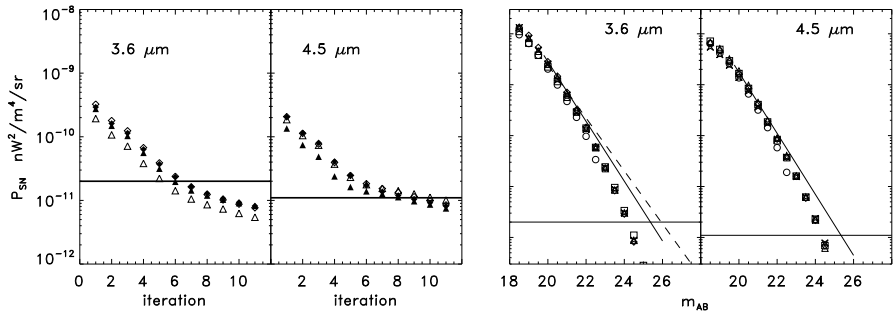}
% \vspace*{-1.0 cm}
 \caption{{\it Left}: Decrease of the shot noise
power vs the iteration number of the source cleaning. Horizontal
lines show the levels reached in \cite{kamm2}, which are a factor
of $\sim 2$ below those in \cite{kamm1}. Four sets of symbols
correspond to the four GOODS fields. Horizontal lines mark the
shot noise levels of KAMM2. {\it Right}: Shot noise power $P_{\rm
SN}$ estimated by integrating the counts. Symbols identical to
those on the left denote the four GOODS fields; open circles
correspond to counts for the QS1700 field. Solid line shows
$P_{\rm SN}$ according to the fit to IRAC counts of \cite{fazio}
used in KAMM1.}
   \label{fig2}
\end{center}
\end{figure}

The extragalactic signal is made of two components: 1) shot noise
from the remaining faint galaxies (shown with dotted lines in Fig.
\ref{fig1}), and 2) on arcminute scales the fluctuations are
produced by clustering of the emitters. It is important to
emphasize that as fainter foreground galaxies are removed so that
the remaining shot noise is reduced the details of the
fluctuations change. This is due to the varying contribution from
the remaining foreground galaxies. The large-scale part of the
fluctuations remains as the foreground sources are removed down to
the lowest shot-noise levels. Left panels of Fig. \ref{fig2} show
the decrease in the shot-noise power, $P_{\rm SN}$, as
progressively higher iterations in the source removal are reached.
The final shot noise reached by us with the GOODS data is shown
with horizontal lines and is a factor of $\sim 2$ lower than in
the QSO 1700 data (KAMM1). The right panels of Fig. \ref{fig2}
compare the final shot noise with that produced by the observed
galaxy counts (which at these wavelengths are confusion limited at
$m_{\rm AB} > 21-22$ for IRAC beam). The figure shows that galaxy
removal is efficient to $m_{\rm AB} > 26-27$ and the signal in
Fig. \ref{fig1} comes from very faint sources.

GOODS fields have also been observed at optical wavelengths with
the {\it Hubble} ACS instruments reaching source detection levels
fainter than 28 AB mag. This allowed us to further test the origin
of the source-subtracted CIB fluctuations. If the latter come from
local populations there should be a strong correlation between the
source-subtracted IRAC maps and the ACS sources. Conversely, there
should be no such correlations if the CIB signal arises at at
epochs where the Lyman break (at rest $\sim 0.1\mu$m) gets shifted
passed the longest ACS z-band at $\simeq 0.9\mu$m. To test for
this in KAMM4 we have constructed synthetic maps, overlapping with
the GOODS fields, using sources in the ACS B, V, i, z bands from
the ACS sources catalog of \cite{goods}. These maps were then
convolved with the IRAC 3.6 and 4.5 $\mu$m beams. Finally, we
applied the clipping mask from the IRAC maps and computed the
fluctuations spectrum produced by the ACS sources and their
correlations with the IRAC-based maps.

The fluctuations in the diffuse light produced by the ACS galaxies
are shown in the right panels of Fig. \ref{fig1}. The contrast
between the spectrum of the source-subtracted CIB fluctuations and
those produced by the optical galaxies is obvious. The former has
the power spectrum such that the fluctuations are flat to slowly
rising with increasing angular scale, whereas the latter have
power spectrum with fluctuation amplitude decreasing with
increasing scale in agreement with CIB measurements from deep
2MASS data arising from galaxies at $z\sim$1-2 (Kashlinsky et al
2002). This by itself shows that the populations producing the
source-subtracted CIB fluctuations are not in the ACS source
catalog.

\begin{figure}[ht]
% \vspace*{-2.0 cm}
\begin{center}
 \includegraphics[width=5in]{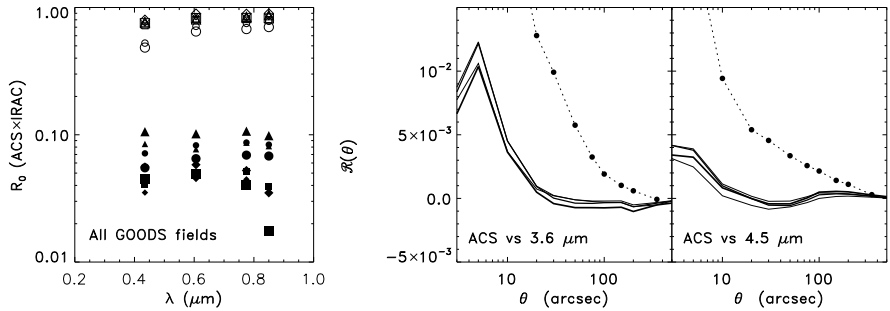}
% \vspace*{-1.0 cm}
 \caption{{\it Left}: Correlation coefficient
between clipped/masked ACS and KAMM data. Large and small symbols
correspond to the IRAC Ch 1 and Ch 2; the four sets of symbols
correspond to the four GOODS fields. Open symbols correspond to
correlations with the maps of the removed sources and filled
symbols with the residual KAMM maps which contain the fluctuations
shown in Fig. 1.  {\it Right}: Solid lines show the dimensionless
correlation function between the diffuse light in the ACS and KAMM
maps for $B,V,i,z$-bands in order of increasing thickness. Dotted
line shows the dimensionless correlation function of the KAMM
maps, $C_{\rm KAMM}(\theta)/\sigma_{\rm KAMM}^2$, which remains
positive out to $\sim 100^{\prime\prime}$ and is better viewed in
log-log plots as in Fig. SI-4 of KAMM1.}
   \label{fig3}
\end{center}
\end{figure}

More importantly, Fig. \ref{fig3} shows that the correlations
between the ACS galaxies and the source-subtracted CIB maps are
very small and on arcminute scales are within the statistical
noise. Thus, at most, the remaining ACS sources contribute to the
shot-noise levels in the residual KAMM maps, but not to the large
scale correlations. At the same time, there are excellent
correlations (shown with open symbols) between the ACS source maps
and the sources {\it removed} by KAMM prior to computing the
remaining CIB fluctuations.

The excess source-subtracted CIB fluctuation on arcminute scales
in the 3.6 $\mu$m channel is $\sim 0.1$ nW/m$^2$/sr; KAMM measure
a similar amplitude in the longer IRAC bands indicating that the
energy spectrum of the arcminute scale fluctuations is flat to
slowly rising with increasing wavelength at least over the IRAC
range of wavelengths. This is illustrated with the solid line in
the left panels of Fig. \ref{fig1}.

\section{Cosmological implications}

Any interpretation of the KAMM results must reproduce three major
aspects:

$\bullet$ The sources in the KAMM data were removed to a certain
(faint) flux limit, so the CIB fluctuations arise in populations
with magnitudes fainter than the corresponding magnitude limit,
$m_{\rm lim}$. Furthermore, these sources are not present among
the optical ACS galaxies as demonstrated by the absence of
correlations between these galaxies and the IRAC source-subtracted
CIB maps.

$\bullet$ These sources must reproduce the excess CIB fluctuations
by KAMM on scales $> 0.5^\prime$.  They must also reproduce the
measured spectrum of the CIB fluctuations, which is different from
the observed clustering pattern of ordinary galaxies at
intermediate $z$.

$\bullet$ Lastly, the populations fainter than the above magnitude
limit must account not only for the correlated part of the CIB,
but - equally important - they must reproduce the (low) shot-noise
component of the KAMM signal, which dominates the power at
$<$0.5$^\prime$. The discussion below by-and-large follows KAMM3:

1) {\bf Magnitude limits and epochs}. Since the ACS galaxies do
not contribute to the source-subtracted CIB fluctuations, the
latter must arise at $z>7$ as is required by the Lyman break at
rest $\sim 0.1 \mu$m getting redshifted past the ACS $z$-band of
peak wavelength $\simeq 0.9 \mu$m. This would place the sources
producing the KAMM signal within the first 0.7 Gyr. If the KAMM
signal were to originate in lower $z$ galaxies which escaped the
ACS GOODS source catalog because they are below the catalog flux
threshold, they would have to be extremely low-luminosity systems
($< 2\times 10^7 h^{-2}L_\odot$ at $z$=1) and these galaxies would
also have to cluster very differently from their ACS counterparts.

 2) {\bf Clustering component}. Solid lines in Fig. \ref{fig1}
show the expected CIB fluctuations from sources with the (biased)
concordance $\Lambda$CDM power spectrum at $z>5$. The fit is
reasonably good making such sources a plausible candidate for
producing the observed signal. At the same time, the observed
galaxy populations out to $z\sim 2-3$ cluster very differently.
Thus any model attempting to assign the KAMM signal to more recent
sources will have to account for this {\it observed} difference in
the clustering patterns.

3) {\bf Net CIB levels from the new sources}: The angle of
$1^\prime$ in the concordance cosmology subtends comoving scales
of 2.2-3 Mpc at 5$\leq z \leq$20. For $\Lambda$CDM density fields
with reasonable biasing one can reach relative arcminute-scale
fluctuations of $\sim$5-10\% meaning that the net CIB from sources
contributing to the KAMM signal at 3.6 $\mu$m is at least 1-2
nW/m$^2$/sr, which is well within the uncertainties of the recent
CIB measurements of Thompson et al (2007).

4) {\bf Shot noise constraints}. The amplitude of the shot-noise
power gives a particularly strong indication of the epochs of the
sources contributing to the KAMM signal. This can be seen from the
expressions for the shot noise (Kashlinsky 2005a):
\begin{equation}
P_{\rm SN} = \int_{m_{\rm AB}>m_{\rm lim}}f(m) dF(m) \equiv
f(\bar{m}) F_{\rm tot}(m_{\rm AB} > m_{\rm lim})\label{shotnoise}
\end{equation}
where $f(m)$ is the flux in Jy of a source of magnitude $m$ and
$F_{\rm tot}(m_{\rm AB} > m_{\rm lim})$ is the net CIB flux
produced by the remaining sources. Above it was shown that the
sources contributing to the fluctuations must have CIB flux
greater than a few nW/m$^2$/sr and combining this with the values
for $P_{\rm SN}\sim 10^{-11}$nW$^2$/m$^4$/sr, reached in the KAMM2
analysis, leads via eq. \ref{shotnoise} to these sources having
typical magnitudes $m_{\rm AB} < 29-30$ or individual fluxes $< 4$
nJy. {\it Such faint sources are expected to lie at very high
$z$.}

5) {\bf Mass/light ratio of the new populations}: This information
on the nature of the populations responsible for these CIB
fluctuations, can be obtained from the fact that the significant
amount of flux ($>$1-2 nW/m$^2$/sr) required to explain the
amplitude of the fluctuations must be produced within the short
time available at these high $z$ (cosmic times $<$0.5-1 Gyr). The
implied comoving luminosity density associated with these
populations is related to the fraction of baryons locked in these
objects with the additional assumption of their $\Gamma\equiv
M/L$. The smaller the value of $\Gamma$, the fewer baryons are
required to explain the CIB fluctuations detected in the KAMM
studies. It turns out that in order not to exceed the baryon
fraction observed in stars, the populations producing these CIB
fluctuations had to have $\Gamma$ much less than the solar value,
typical of the present-day populations (KAMM3). This is consistent
with the general expectations of the first stars being very
massive.

6) {\bf Resolving the new sources}: In order to directly detect
the faint sources responsible for the CIB fluctuations with fluxes
below a few nJy, their individual flux must exceed the confusion
limit. If such sources were to contribute to the CIB required by
KAMM data, at 3.6 and 4.5 $\mu$m they had to have the average
surface density of $\bar{n} \sim F_{\rm CIB}^2/P_{\rm SN} \sim 5
\; {\rm arcsec}^{-2}$. In order to avoid the confusion limit and
resolve these sources individually at, say, 5-sigma level
($\alpha=5$) one would need a beam of the area $\omega_{\rm beam}
\leq \alpha^{-2}/\bar{n} \sim 5\times 10^{-3}{\rm arcsec}^2$ or of
circular radius below $\sim$0.04 arcsec. This is clearly not in
the realm of the currently operated instruments, but the {\it
JWST} could be able to resolve these objects given its sensitivity
and resolution.

{\bf Acknowledgements} I thank my collaborators, Rick Arendt, John
Mather and Harvey Moseley for many contributions to the KAMM
results, and the NSF AST-0406587 grant for support.

\end{document}